\documentclass[conference]{IEEEtran}
\makeatletter
\def\ps@headings{%
\def\@oddhead{\mbox{}\scriptsize\rightmark \hfil \thepage}%
\def\@evenhead{\scriptsize\thepage \hfil \leftmark\mbox{}}%
\def\@oddfoot{}%
\def\@evenfoot{}}
\makeatother
\IEEEoverridecommandlockouts
\usepackage{cite}
\usepackage{amsmath,amssymb,amsfonts}
\usepackage{algorithmic}
\usepackage{graphicx}
\usepackage{textcomp}
\usepackage{subcaption}
\usepackage{xcolor}
\def\BibTeX{{\rm B\kern-.05em{\sc i\kern-.025em b}\kern-.08em
    T\kern-.1667em\lower.7ex\hbox{E}\kern-.125emX}}
\begin{document}

\title{More or Less? Predict the Social Influence of Malicious URLs on Social Media\\
}

\author{\IEEEauthorblockN{Chun-Ming Lai\IEEEauthorrefmark{1},
Xiaoyun Wang\IEEEauthorrefmark{1},
Jon W. Chapman\IEEEauthorrefmark{1}, 
Yu-Cheng Lin\IEEEauthorrefmark{1},\\
Yu-Chung Ho\IEEEauthorrefmark{1},
S. Felix Wu\IEEEauthorrefmark{1},
Patrick McDaniel\IEEEauthorrefmark{2} and
Hasan Cam\IEEEauthorrefmark{3}
}
\IEEEauthorblockA{\IEEEauthorrefmark{1}
	University of California, Davis\\
\{cmlai,xiywang,jwchapman,ycjlin,ycaho,sfwu\}@ucdavis.edu
}
\IEEEauthorblockA{\IEEEauthorrefmark{2}
	Pennsylvania State University\\
	mcdaniel@cse.psu.edu
}

\IEEEauthorblockA{\IEEEauthorrefmark{3}
	U.S. Army Research Laboratory\\
	hasan.cam.civ@mail.mil
}
}
\maketitle

\begin{abstract}
Users of Online Social Networks (OSNs) interact with each other more than ever. In the context of a public discussion group, people receive, read, and write comments in response to articles and postings. In the absence of access control mechanisms, OSNs are a great environment for attackers to influence others, from spreading phishing URLs, to posting fake news. Moreover, OSN user behavior can be predicted by social science concepts which  include conformity and the bandwagon effect.
In this paper, we show how social recommendation systems affect the occurrence of malicious URLs on Facebook. We exploit temporal features to build a prediction framework, having greater than 75\% accuracy, to predict whether the following group users' behavior will increase or not. Included in this work, we demarcate classes of URLs, including those malicious URLs classified as creating critical damage, as well as those of a lesser nature which only inflict light damage such as aggressive commercial advertisements and spam content. It is our hope that the data and analyses in this paper provide a better understanding of OSN user reactions to different categories of malicious URLs, thereby providing a way to mitigate the influence of these malicious URL attacks.
\end{abstract}

\section{Introduction}
The attack vectors that users of Online Social Networks (OSNs) face have been evolving as the various bad actors learn to manipulate this new aspect of the cyber landscape. One of these newly evolving attack vectors is the news creation and dissemination cycle. Traditionally the structure for media dissemination was a top-down arrangement, news was typically published by well-trained reporters and edited by a skilled team. In this fashion, these professionals acted as a gatekeeper of sorts, ensuring higher journalistic integrity and correctness of the news. In contrast, news is now being created for use and spread on Online Social Networks by users. This opens the door to the news generation process, and the associated URLS, to be utilized as nascent attack vectors.

With the modern structuring of this information creation and consumption process, social recommendation systems play an increasingly critical role as they determine which users will see exactly what information based on the characteristics of the individual users as well as specific features from the articles. Unfortunately, inappropriate information will diffuse on user-generated content platforms much more readily than traditional media, attributable to two primary factors: (1)
The connectivity of social media makes information diffusion deeper and wider. (2) Users may wittingly or unwittingly boost inappropriate dissemination cascades with their own comments on the articles. 

The major research efforts in the area of OSN security are concerned with detecting malicious accounts rather than normal user accounts. However, attackers can exploit either newly created fake or existing latent compromised accounts to avoid state-of-the-art defense schemes since most are based on verified attacker behavior and trained by machine learning algorithms either by lexical features or accounts characteristics. Relatively less work has been done to measure and consider the influence of the actual malicious content. Therefore, our motivation is based on two research questions: RQ1. For discussion threads having clearly malicious content, do they have a larger cascade size when compared to other discussion threads that do not contain malicious content? RQ2. In a discussion thread, is there a significant influence between the prior and the latter comments surrounding a malicious comment?
Specifically, we are wondering will audiences change their behavior when they see a malicious comment which is being \emph{promoted} by a social recommendation system.

In this paper, we design two experiments trying to answer the above two questions. For cascade size between target and non-target post threads, we evaluate with the bandwagon effect experiment. 
The findings indicate that they both basically follow the same cascade model. However, their final cascade size are extremely different for at least two reasons: (1) Users reactions and (2) Social Recommendation Design. Afterwards, we turned our attention to the fine-grained influence of a user-generated comment. We define an Influence Ratio (IR) for every comment to evaluate its influence based on the ratio over its upcoming activities and its preceding activities. Our framework achieves more than 75\% accuracy on both critical and light damage URLs in predicting the upcoming activity increase or decrease. 

\begin{figure}[]
	\centering
	\includegraphics[width=.6\columnwidth]{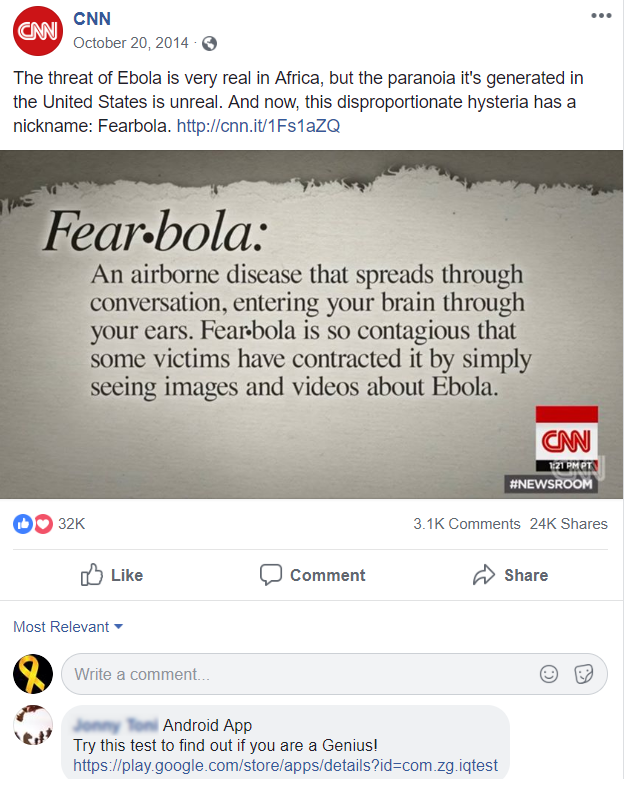}
	\caption{Android apps download links posted to an article about the threat of Ebola.}
	\label{light}
\end{figure}

\begin{figure}[]
	\centering
	\includegraphics[width=.6\columnwidth]{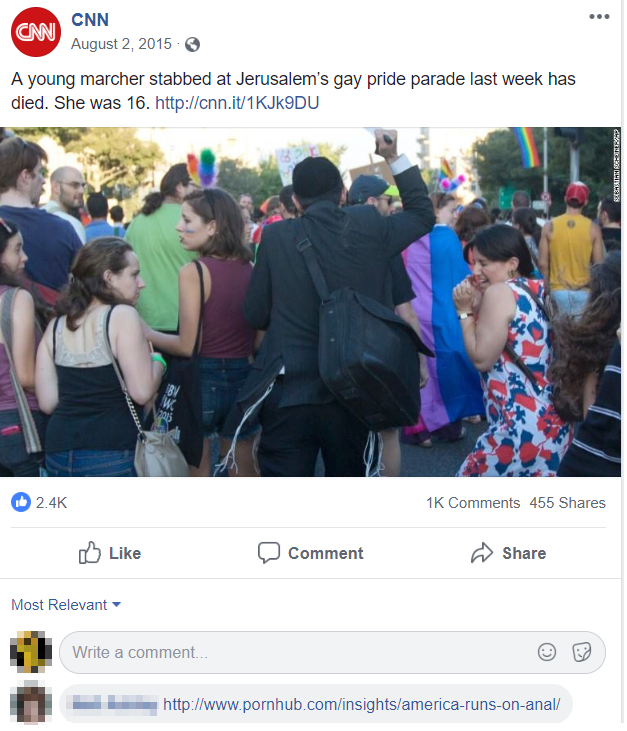}
	\caption{Pornography URL posted to comment stream of article about an LGBT pride parade}
	\label{porn}
\end{figure}

Our results also indicate the relative position, in context of the chronological positioning, of a comment plays a critical role in contributing to the influence that the comment wields. For example, Figure \ref{light} \footnote{post\_id = 10152998395961509} shows a labeled advertisement URL occurred in the very late stage (2802 out of 2844 total comments) while Figure \footnote{post\_id = 10153908916401509} \ref{porn} shows a labeled pornography URL occurred near the middle of the comment chronology (470th comment of 942 total). We found that with regards to position, critical and those lesser threat level URLs are presenting at different times in the discussion --- the light threat URLs tend to be posted later chronologically while critical threat URLs tend to be in the middle of a post's timeline. This phenomenon is borne out very obviously in the CNN public page, however this presentation is not as dramatic on the FOX News. There are at least two reasons for the chronological disparity between the different threat level URLs: (1) Users tend to leave the discussion when they feel there was an obvious ad posted, like Figure \ref{light}. (2) Compared to the lower level threats, attackers who spread critical malicious URLs act in a more strategic manner --- they choose the most opportune timing to achieve the greatest amount of influence.

The rest of this paper is organized as follows: Section \ref{section2} 
illustrates how Facebook and the Social Recommendation Systems work. In Section \ref{section3} we define necessary terms and provide a detailed description for our dataset. The bandwagon effect cross-validation is described in Section \ref{section4}. We define and predict the Influence Ratio in Section \ref{section5}. Related Work and our Conclusion are given in Section \ref{section6} and Section \ref{section7}, respectively.

\section{Facebook and Social Recommendation System}

\label{section2}
In this section, we introduce one of the most popular online services of social media --- Facebook, from it's humble beginnings as a sort of simple digitized social yearbook limited to only certain universities to a worldwide  incredibly complex and multi-functional platform. We also describe the information consumption process between Facebook and OSNs Users.

\subsection{Facebook public pages}
Facebook was launched in 2004, initially providing a platform for students to search for people at the same school and look up friends of friends. Users updated personal information, likes, dislikes, as well as current activities. While doing this they also kept track of what others were doing, and used Facebook to check the relationship availability of anyone they might meet and were interested in romantically. \footnote{https://www.eonline.com/news/736769/this-is-how-facebook-has-changed-over-the-past-12-years} 
As Facebook grew quickly, users were not satisfied with merely following the personal status of their close friends on the network. Furthermore, users demonstrated an interest in public affairs and news. 
For this reason, the public pages on Facebook were created, and have become places where users receive news and information \emph{selected} and \emph{promoted} by news feeds, which are constantly updating lists of stories in the middle of one's homepage, including stories regarding (1) friends' activities on Facebook. (2) articles from pages where an user is interested (Liked or Followed). (3) articles that your friends like or comment on from people you are not friends with. (4) Advertisements from sponsoring companies and organizations. \footnote{https://www.facebook.com/help/327131014036297/}. With these new media publication venues on Facebook, users interact with strangers on various public pages --- discussing news published by commercial media companies, announcements by public figures, sharing movie reviews, gossiping about an actor, or criticizing the poor performance of a particular sports teams. 
According to \cite{hong2018profiling}, there are more than 38,831,367 public pages covering multiple topics including Brands \& Products, Local Business \& Places and Companies \& Organizations.  \footnote{https://www.facebook.com/pages/create}.  

\subsection{Social Recommendation System}
\begin{figure*}[]
	\centering
	\includegraphics[scale=0.6]{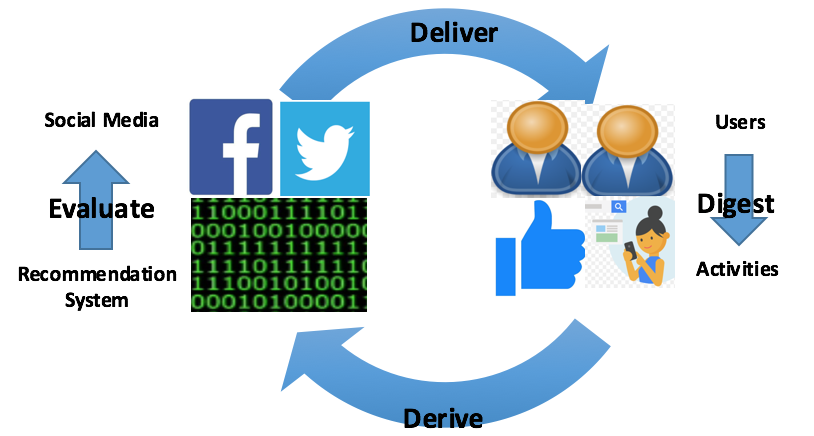}
	\caption{The framework of users activities regarding Online Social Media}
	\label{recommend}
\end{figure*}

Most highly trafficked Online Social Media sites contain some variation of a dynamic social recommendation system \cite{ricci2011introduction}. It is a continuous process cycle, which includes two entities : \textbf{Social Computing Platform} and \textbf{Active Users} and the four processes shown in Figure \ref{recommend}. Here we explain the processes in more detail.

\begin{enumerate}
	\item Deliver: Large-scale and user-generated data have been disseminated on OSNs. However, only appropriate information is delivered to corresponding audiences.
	\item Digest: When users see the news, they will be able to have a chance to join the discussion by actively typing their opinion, less actively clicking reactions or passively doing nothing.
	\item Derive and Evaluate: Recommendation systems will collect a large amount of user interaction data and modify the algorithm to better attract attention from users (mostly because of attention economy \cite{davenport2001attention}). The evaluation step gives a chance for Facebook to modify the social algorithms to deliver more appropriate (Step 1.) information to users. The primary concern for Facebook is to maximize clicks on advertisements. This is primarily accomplished by maximizing time spent on Facebook by the user. 
\end{enumerate}

Attackers logically attempt to maximize the influence they wield for every malicious campaign. In effect, having more people see their malicious content, click it, interact with it, or trust it. With the application of behavioral targeting, we believe the bad actors spread URLs that will be more relevant to audiences, whose patterns could be collected from data mining or speculation. For example, the collection of bad actors involved with spreading fake news tend to chronologically target the planting of their fake news as well as topographically target the best locations to plant fake news (eg. Politics-related Facebook pages or articles). While other bad actors run accounts that are hired by commercial enterprises that have a more limited scope and primarily care about their business. The common thread is they all make use of the social recommendation system. 
We have seen that Social Recommendation System design actually increases the damage of malicious URLs since it offers a way for attackers to spread harmful content at the right place and at the right time. Vosoughi et al. indicates that false news is more novel than true news and humans are more likely to share novel information online \cite{vosoughi2018spread}. Therefore, the social recommendation system will boost the "rich get richer" effect. 

\section{Data Description and Labeling }
\label{section3}

In this section, we define the necessary terminology used in this paper. After providing a high level overview of the discussion groups dataset, we show how we label filtered URLs into different categories.

\subsection{Terminology}
We use the following terminology to describe the concepts in our work more exactly:
\begin{itemize}
	\item page: a public discussion group. In this study, we only consider two main media pages: CNN ($page\_id = 5550296508$) and FOX News ($page_id = 15704546335$) . 
	\item Original Post: an article on a Facebook discussion group.
	\item Comments: text written in response to an original post.
	\item Reaction: Emoji to comments or original posts, including "Like", "Love", "Haha", "Love", "Sad" and "Angry".
	\item Post thread: the original post and all corresponding user activities (Comments and corresponding Reactions), ordered by their timestamps.
	\item Target post thread: post threads which have at least one comment embedded with malicious URL(s).
	\item Non-Target Posts: post threads which have no embedded malicious URLs.
	\item Time Series TS: $TS_{created}$ indicates a time period $j$ following the time of the original post, measured in minutes. $TS_{final}$ refers to the precise time 1 hour after the original post was created (i.e., $final$ = 60).
	\item Number of comments $N_{comment}(post, TS_i)$: the number of post comments collected at $TS_i$
	\item Accumulated number of participants $AccN_{comment}(post,TS_i)$: the number of post comments between $TS_i$ and $TS_{i-1}$
	
\end{itemize}

\subsection{Crawled Dataset}
To this end, our data was cataloged with the use of an open source social crawler called SINCERE \footnote{https://github.com/dslfaithdev/SocialCrawler}(Social Interactive Networking and Conversation Entropy Ranking Engine), which has been created and refined by our research group over several years. We employed it from 2014 to 2016 to collect post threads on both CNN and FOX News public pages to see the difference between left-wing and right-wing discussion. Detail information was stored, including timestamps of each comment, Facebook account identification numbers, as well as the raw text of comments and articles. In total we have 48,087 posts, 88,834,886 comments and 189,460,056 reactions for both pages. We describe the full dataset in Table \ref{data_desc}. 

\begin{table*}
	\centering
	\caption{Data Description}
	\label{data_desc}
	\begin{tabular}{c|c|c|c|c|c} \hline
		Page Name & Total Posts &  Total Comments & Comments with URLs &  Total Reactions\\ \hline
		CNN & 20922 & 11,882,590&  412,001 (3.47\%) & 24,174,160\\
		FOX News & 27165 & 76,952,296 & 1,026,525 (1.33\%)& 165,285,896\\
		\hline\end{tabular}
\end{table*}

\subsection{Labeling URLs}
Typically, a URL contains three parts: (1) a protocol, (2) a host name and a file name. In this paper, we focus on URLs which use HTTP and HTTPS protocols. Moreover, we focus on the host name itself. We first use a well-known Whitelist \emph{{'facebook.com', 'youtube.com', 'twitter', 'on.fb.me', 'en.wikipedia', 'huffingtonpost.com', 'foxnews.com', 'cnn.com', 'google.com', 'bbc.co.uk', 'nytimes.com', 'washingtonpost.com'}} to do a first-step filter.
 We then employ the daily-updated Shalla Blacklist service \cite{Shalla}, which is a collection of URL lists grouped into several categories intended for the usage with URL filters, to label and trace the behavior of URL influence. Note that we do not assume all URLs filtered by Shalla are completely malicious. Among the 74 categories listed, we manually divided targeted URLs into two classes: Light and Critical, the explanation of each category is follows: 
\begin{itemize}
	\item \textbf{Light}:
		\begin{itemize}
		    \item Advertising: Includes sites offering banners and advertising companies.
			\item Shopping: Sites offering online shopping and price comparisons.
			\item Gamble : Poker, Casino, Bingo and other chance games as well as betting sites.
			\item Porn: Sites with sexual content.
		\end{itemize}
	\item \textbf{Critical}:
		\begin{itemize}		
			\item Download: This covers mostly file-sharing, p2p, torrent sites and drive-by-downloads. 
			\item Hacking: Sites with information and discussions about security weaknesses and how to exploit them.
			\item Spyware: Sites that try to actively install software (or lure the user to do so) in order to spy on user behavior. This also includes trojan and phishing sites.
			\item Aggressive: Sites of obvious aggressive content. Includes hate speech and racism.
			\item Drugs: Sites offering drugs or explaining how to make drugs (legal and non legal). 
			\item Weapons: Sites offering weapons or accessories for weapons
			\item Violence: Sites about killing or harming people or animals.
		\end{itemize}
\end{itemize}
We classify others as Benign if they are not in the Whitelist, Light, or Critical classes. The detailed number of each category is listed in Table \ref{urldata_desc}.

\begin{table*}
	\centering
	\caption{URL Data Description}
	\label{urldata_desc}
	\begin{tabular}{c|c|c|c|c} \hline
		Page Name & URL in WhiteList &  URL in  Light &  URL in Critical & URL in Benign\\ \hline
		CNN & 194,372 & 3,762 &  636 & 213,231\\
		FOX News & 503,480 & 8,125 & 1,571 & 513,349\\
		\hline\end{tabular}
\end{table*}

\section{Post level Influence}
Heterogeneous posts are updated and refreshed at tremendous speed and include Videos, photos with attractive headlines, assorted topics such as international affairs, elections or entertainment. We are interested in why malicious URLs often occur in only some post threads. In this section, we applied the model proposed by Wang et.al \cite{wang2015bandwagon} to gain insight into how those malicious URLs may influence the growth of the conversation. 

\label{section4}
\subsection{Bandwagon Effect and Attacker cost}
A phenomenon known as the bandwagon effect, which explains how individuals will agree with a larger group of people who they may not normally agree with, but do so in order to feel a part of a group, --- individuals are more likely to be with those sub-groups that share similar thoughts but feel uncomfortable in the presence of minority groups that have different ideas \cite{allen1965situational}. Many voting behaviors are related to this effect, voters may or may not follow their own conscience to make a voting decision but may just follow the majority opinion \cite{van2015off}. 

From the result obtained by \cite{lai2017attacking}, when considering the posts targeted with a comment that includes a malicious URL, we see the most commonly attacked articles tend to generate large amounts of discussion. Moreover, targets may be those \emph{suggested} by the Facebook social recommendation system to the attacker.

The following example indicates how a large number of majority opinions might be identified from a Facebook discussion. Assume three posts --- $post_A$, $post_B$ and $post_C$ --- have been posted on a public page at around the same time, the original posters' identities are irrelevant. Also assume there exist three different users --- $user_A$ visits the page and browses all three posts, has no signals from others for making a decision to engage with the post, the user subsequently decides to only comment on $post_B$ because it was subjectively the most interesting one to them. Five minutes later, $user_B$ visits the same page and sees that only $post_B$ has a comment. This user then checks $post_B$ first and decides to add their own reply, either in response to the original post or $user_A$. Note that until now there are no comments on either $post_A$ or $post_C$. 
Some short time later, $user_c$ checks the page and finds that $post_B$ has more than 10 comments, while $post_A$ and $post_C$ still have 0 comments. He then decides to add a comment to $post_B$ since $post_B$ is the first post that is pushed to the user by Facebook because, at that moment, $post_B$ has a relatively larger share of public attention compared to $post_A$ or $post_C$. This is an example of the information cascade phenomenon first proposed by Bikhchandani, Hirshleifer and Welch \cite{bikhchandani1992theory}, and most social media recommendation systems intensify this phenomenon --- information and user activities are automatically selected by algorithm, though most users do not realize this when participating in OSN discussion groups.

\subsection{Prediction Model and Evaluation Method}
In order to differentiate the information cascade model between target and non-target post threads, we will describe a system designed to use the time series and number of current comments to predict how many new users are likely to participate in each respective thread. 
The Discussion Atmosphere Vector (DAV) \cite{lai2017attacking} defined in our previous work used the definition of accumulated number of participants given in Section \ref{section3}, using 5 minutes for the $i$ value and 2 hours for the $t_final$ value.
\begin{equation}
\begin{aligned}
DAV(Post)_{t_n} = [AccNcomment(Post,t_1),\\
AccNcomment(Post,t_2), ...,  \\ 
AccNcomment(Post,t_n)]\nonumber
\end{aligned}
\end{equation}

In the bandwagon effect model proposed by Wang \emph{et. al} \cite{wang2015bandwagon}, the numbers of comments with respect to each time window after a post has been created can be used to build matrices for each public page $G$ to predict the final number of comments by machine learning and statistical methods. 
In other words, two post threads $post_A$ and $post_B$ are likely to have the same scale of cascades if in each timestamp $i$ such that:  
\begin{equation}
DAV(Post_A )_{t_i} \approx DAV(Post_B)_{t_i}\nonumber
\end{equation}

We then defined a distribution matrix $D$, with each element $D_{ij}$ representing a set of posts $post \in G$, including the final number of comments we crawled $N_{comment}(post, TS_{final})$ and aggregate number of comments $j (j = N_{comment(post,TS_i)})$ at time $i$.

\begin{equation}
\begin{aligned}
D_{ij}(G) = \{N_{comment}(post,TS_{final}(post)) | \\
j = N_{comment}(post,TS_i(post)); \forall post \in G\}
\end{aligned}
\end{equation}

Based on the distribution matrix $D$, we used a bootstrapping method \cite{efron1994introduction} to construct prediction matrix $M$.
\begin{equation}
\begin{aligned}
M_{ij} = Bootstrapping(D_{ij})
\end{aligned}
\end{equation}
The matrix $M$ is used to create a prediction function $F_{predict}$ that collects two inputs from any new post thread: observed time series $TS_{ob}(post)$ and corresponding feature $N_{comment}(post,TS_{ob})$. According to $M$, we obtain the result using the following equation:
\begin{equation}
\begin{aligned}
F_{predict}(TS_{ob}(post),N_{comment}(post,TS_{ob}) = \\
M_{TS_{ob}(post, N_{comment}(post,TS_{ob}))} 
\end{aligned}
\end{equation}

\begin{table}
	\centering
	\caption{Bandwagon Effect Cross Validation -- Target to Non-Target, Observed time = 120 minutes}
	\label{BEI_TtoN_result}
	\begin{tabular}{c|c|c} \hline
		Page Name & Precision / Predictable (\%) & Predictable / All (\%) \\ \hline
		CNN & 15,633/15,866 (97\%) & 15,866/15,869 (99\%) \\ 
		FOX News & 17,338/17448 (99\%) &17,448/17,453 (99\%) \\		
		\hline\end{tabular}
\end{table}

\begin{table}
	\centering
	\caption{Bandwagon Effect Cross Validation -- Non-Target to Target, Observed time = 120 minutes}
	\label{BEI_NtoT_result}
	\begin{tabular}{c|c|c} \hline
		Page Name & Precision / Predictable (\%) & Predictable / All (\%) \\ \hline
		CNN & 5,013/5,014 (99\%) & 5,014/5,053 (99\%) \\ 
		FOX News & 9,706/9,712 (99\%) &9,712/9,712 (100\%) \\		
		\hline\end{tabular}
\end{table}

\subsection{Result and Discussion} 

Our Bandwagon cross-validation between Target and Non-Target post threads are offered in Tables \ref{BEI_NtoT_result} and \ref{BEI_TtoN_result}. Note that the prediction function can sometimes fail for either of two reasons: insufficient features for a post to match M in the testing data or insufficient existing posts within the training data. In our experiment, we have enough training data, so unpredictable posts $\approx 1\%$ for both pages are posts which do not have enough activities for M to predict the final size of cascade.

Basically there are no obvious differences regarding the first two hours activities with respect to final number of comments between Target and Non-Target post threads. This suggests that malicious URLs did not affect the life cycle of post threads --- people still engaged the target post threads, under the threat of malicious URLs. On the other hand, our result also indicate that the Facebook social recommendation system continued to \emph{deliver} post threads which have malicious URLs to audiences, similar to the way it treats normal post threads. 

Recall that the Bootstapping method in this experiment is only providing the lower bound. For example,if $M_{5,5} = 100$, this means any post satisfies 5 comments in the first five minutes, it would have 100 comments or more. However, 200 and 2000 are both greater than 100, but the scale are not the same. 
In order to consider the final cascade of comments between targets and non-targets, we also conduct two-sample Kolmogorov-Smirnov (KS) tests to compare the distributions of the final number of comments between those two sets. The results are shown in Table \ref{KSCNN} and \ref{KSFOX}. In general, for both the FOX News and CNN pages, the final number of comments of Target post threads is obviously greater than Non-Target ones. There are two main reasons: (1) Attackers are \emph{led} by the Facebook Social Recommendation Systems. In other words, their targets are not chosen by themselves but mostly by Social Algorithms. (2) Normal users tended to react more than usual because of those malicious URLs --- novel information would ignite interest to join a discussion. We also noticed that FOX News attracts more people to join the discussion, rather than CNN (about 4 times the mean of the number of comments per post threads).

\begin{table}
	\centering
	\caption{CNN Statistics on the cascade size. \\
	KS-test for target and non-targets: $D=0.068, p \approx 0.0$}
		\label{KSCNN}
		\begin{tabular}{c c c c c c } \hline
		      & 	N  & Mean &  SE & Min & Max \\ \hline \hline
	Target & 5,053 & 1,010 & 1,558 & 14 & 39,929 \\ 
	Non-Target & 15,869 & 427 &  698 & 1 & 35,591 
	     \end{tabular}	
\end{table}

\begin{table}
	\centering
	\caption{FOX Statistics on the cascade size. \\
		KS-test for target and non-targets: $D=0.066, p \approx 0.0$}
	\label{KSFOX}
	\begin{tabular}{c c c c c c } \hline
		& 	N  & Mean &  SE & Min & Max \\ \hline \hline
		Target & 9,712 & 4,712 &11,740 & 24 & 412,621  \\ 
		Non-Target & 17,453 & 1,786 & 3,436  & 1 & 115,669 
	\end{tabular}	
\end{table}

\section{Influence Ratio of a comment}

\label{section5}
Post thread is a basic unit to consider the interaction of users. In previous section we have shown that the final size of cascade (number of comments) and the first two hours activities between target posts and non-targets are very similar, however, the scale are extremely different from KS test. In this section, we turn our attention to the temporal neighbor --- Users interact with a time period on the same post threads, even though they are not mutual friends on Facebook.

\subsection{Preceding and Upcoming Activities}

Consider an original post released by a news media outlet on its public page. This post can be a video, a photo or even just a short paragraph of text. There can be many users activities toward this particular post. Consider an original article ($post$). The corresponding comments are $C_1, C_2,...,C_n$, ordered by their created timestamp ($Time(C_1), Time(C_2), ..., Time(C_n)$). In order to evaluate the influence of a comment $C_i$ with its created time $Time(C_i)$, given a time window $\Delta T$, we define \emph{Influence Ratio (IR)} as the log ratio between all activities which occurred in the previous time window $Time(C_i) - \Delta T$ and the upcoming time window $Time(C_i) + \Delta T$. 
\begin{equation}
\begin{split}
Influence Ratio (C_i, \Delta T) = \\
    log(\frac{count(activities) \in (Time(C_i) +\Delta T)}{count(activities) \in (Time(C_i) -\Delta T)})
\end{split}
\end{equation}
we classify the comment itself in the time period $(Time(C_i) -\Delta T$ to avoid the denominator becoming $zero$. 
Activities include all comments, likes and reactions. If IR is greater than $0$, This means people will be more interested in this post and this comment in the next time slot $Time(C_i) + \Delta T$. 

\subsection{Predict and Evaluate Influence Ratio}

The time differences between two consecutive comments $C_n - C_{n-1}$ vary a lot. For example, 
several studies have shown that post threads would have a \emph{rich get richer} \cite{helsper2017rich} and \emph{bandwagon effect} \cite{lee2016predicting}, which indicates the nearby comments and reactions are critical to interact and influence with each other --- everyone is a potential amplifier. Consider two users $User(C_i)$ and $User(C_j)$ who contribute $C_i$ and $C_j$. If $\mid i-j \mid$ is quite close to 1, they would have a higher chance to interact with each other since the (1) Social recommendation system delivered this post to both users because of their past activities and browsing footprints. (2) They remain online on Social Media at around the same time (This is not always true since we also need to consider the time difference between $Time(C_i)$ and $Time(C_j)$. 
However, they may not be friends with each other but just have overlapping active time on Facebook.
In order to consider the volume of specific time period, we define a function $CountActivity(post,[Time(Begin), Time(After)])$ which refers to all activities (comments, likes, reactions and replies) for $post$ within $Time(Begin)$ to $Time(After)$.
In order to consider the influence and role of a comment in a post thread, given an influenced threshold $\delta T$ and preceding audience number $N_{prev}$, we define Preceding Influenced Vector(PIV) of a comment $C_k$ in a post thread $Post$ as following:

\begin{equation}
\begin{split}
    PIV_i(Post, C_k) = CountActivity(Post, \\ 
     [Time(C_k) - i *\delta T, Time(C_k) - (i*1) \delta T])
\end{split}
\end{equation}

Our goal is to predict IR, the volume of the upcoming time windows. In other words, for any comment $C_k$ in an article $post$, the influence ratio problem predicts whether the upcoming audiences will be greater than the preceding audiences via a classifier --- $c ---> {target,nontarget}$ --- based on one set of $PIV(Post, C_k)$. Hence, for two arbitrary comments $C_m$ of $Post_m$ and $C_n$ of $Post_n$, they will be more likely to have the same trend for the upcoming number of activities such that:

\begin{equation}
PIV(Post_m, C_m) \approx PIV(Post_n, C_n) \nonumber
\end{equation}

We use comments with benign URLs as a training data, trying to predict the IR trends for both light URLs and critical URLs. We set time window $\delta T$ as 1 minute and the number of components of PIV as $60$, which means that for each comment, we assume the time period up to 1 hour will influence the IR.
We then normalized the input PIV to prevent overfitting. As output, we labeled the positive value of IR as \emph{increase} while negative value of IR as \emph{decrease}.
We applied two popular machine learning classifiers (1) Adaboost (2) Gaussian Naive Baynes and from scikit-learn \cite{Pedregosa:2011:SML:1953048.2078195}. 
For Adaboost, we set the number of estimators = 50 and learning rate = 1.
The detailed results are shown in Tables \ref{IR_BenightoLight} and \ref{IR_BenightoCritical}. Overall we can achieve greater than 75\% F1-score on predicting the Influence ratio for both Light and Critical URLs, and the result for Light category is better than Critical category. Moreover, we summarize our findings as follows:
\begin{itemize}
	\item CNN vs. FOX News: There are no obvious difference between CNN and FOX News with regards to predicting IR for the more critical threats versus the lower threat malicious campaigns. We think the reason may be both CNN and FOX News Feeds were controlled by the same social recommendation system. Hence, user activities with respect to temporal features can be predicted with the same amount of ease on either feed. 
	\item Increase vs. Decrease: For most cases, the F1-score on predicting increase is better than decrease (CNN benign to Critical and FOX News both cases). We think it is related to our previous experiment regarding bandwagon effect. We also notice that on the CNN page, the IR of light malicious campaigns tend to decrease, which can either be audiences left the discussion because of the URL or the attacker strategies are inefficient to cause popularity.
	\item Classifiers:  Better results were obtained by Adaboost. For social media data, since PIVs are not independent with respect to one another, Naive Baynes does not work well given the dependent variable constant. In other words, when considering group behaviors on OSNs, we believe that Reinforcement Learning classifier is better than Naive Baynes, which is based on probabilistic classifiers.
\end{itemize}

\begin{table}
	\caption{Influence Ratio Prediction -- Benign to Light, Observed time = 60 minutes, $\delta T$ = 1 minute}
	\label{IR_BenightoLight}
	\begin{tabular}{c|c|c|c|c} \hline
		  & Precision  & Recall & F1-score & Number of Samples \\ \hline
		CNN &  & & & \\ 
		\textbf{Naive Baynes}&  & & & \\
		Decrease & 0.86 & 0.49 & 0.62 & 2,267 \\
		Increase  &  0.53 & 0.88 & 0.66 & 1,495 \\
		avg/total & 0.73 & 0.64 & 0.64 & 3,762 \\ 
		\textbf{Adaboost}  &  & & & \\
		Decrease & 0.84 & 0.80 & 0.82 & 2,267 \\
		Increase  &  0.72 & 0.77 & 0.74 & 1,495 \\
		avg/total & \textbf{0.79} & \textbf{0.79} & \textbf{0.79} & 3,762 \\ 
		\hline
		FOX News &  & & & \\
		\textbf{Naive Baynes}&  & & & \\
		Decrease & 0.69 & 0.42 & 0.52 & 3,638 \\
		Increase  &  0.65 & 0.85 & 0.74 & 4,487 \\
		avg/total & 0.67 & 0.66 & 0.64 & 8,125 \\ 
		\textbf{Adaboost}  &  & & & \\
		Decrease & 0.81 & 0.69 & 0.74 & 3,638 \\
		Increase  &  0.78 & 0.87 & 0.82 & 4,487 \\
		avg/total & \textbf{0.79} & \textbf{0.79} & \textbf{0.79} & 8,125 \\ 
		\hline		
	\end{tabular}
\end{table}

\begin{table}
	\caption{Influence Ratio Prediction -- Benign to Critical, Observed time = 60 minutes, $\delta T$ = 1 minute}
	\label{IR_BenightoCritical}
	\begin{tabular}{c|c|c|c|c} \hline
		& Precision  & Recall & F1-score & Number of Samples \\ \hline
		CNN &  & & & \\ 
		\textbf{Naive Baynes}&  & & & \\
		Decrease & 0.88 & 0.42 & 0.57 & 318 \\
		Increase  &  0.62 & 0.94 & 0.75 & 318\\
		avg/total & 0.75 & 0.68 & 0.66 & 636 \\ 
		\textbf{Adaboost}  &  & & & \\
		Decrease & 0.83 & 0.71 & 0.77 & 318 \\
		Increase  &  0.75 & 0.86 & 0.80 & 318\\
		avg/total & \textbf{0.79} & \textbf{0.79} & \textbf{0.79} & 636 \\ 
		\hline
		FOX News &  & & & \\
		\textbf{Naive Baynes}&  & & & \\
		Decrease & 0.59 & 0.48 & 0.53 & 581 \\
		Increase  &  0.73 & 0.81 & 0.77 & 990 \\
		avg/total & 0.68 & 0.69 & 0.68 & 1,571 \\ 
		\textbf{Adaboost}  &  & & & \\
		Decrease & 0.75 & 0.56 & 0.64 & 581 \\
		Increase  &  0.78 & 0.89 & 0.83 & 990 \\
		avg/total & \textbf{0.77} & \textbf{0.77} & \textbf{0.77} & 1,571 \\ 
		\hline		
	\end{tabular}
\end{table}

\subsection{Life Cycle stage}
\begin{figure*}[]
	\centering
	\hspace*{-1.0cm}
	\begin{subfigure}[]{0.45\textwidth}
		\includegraphics[scale=0.18]{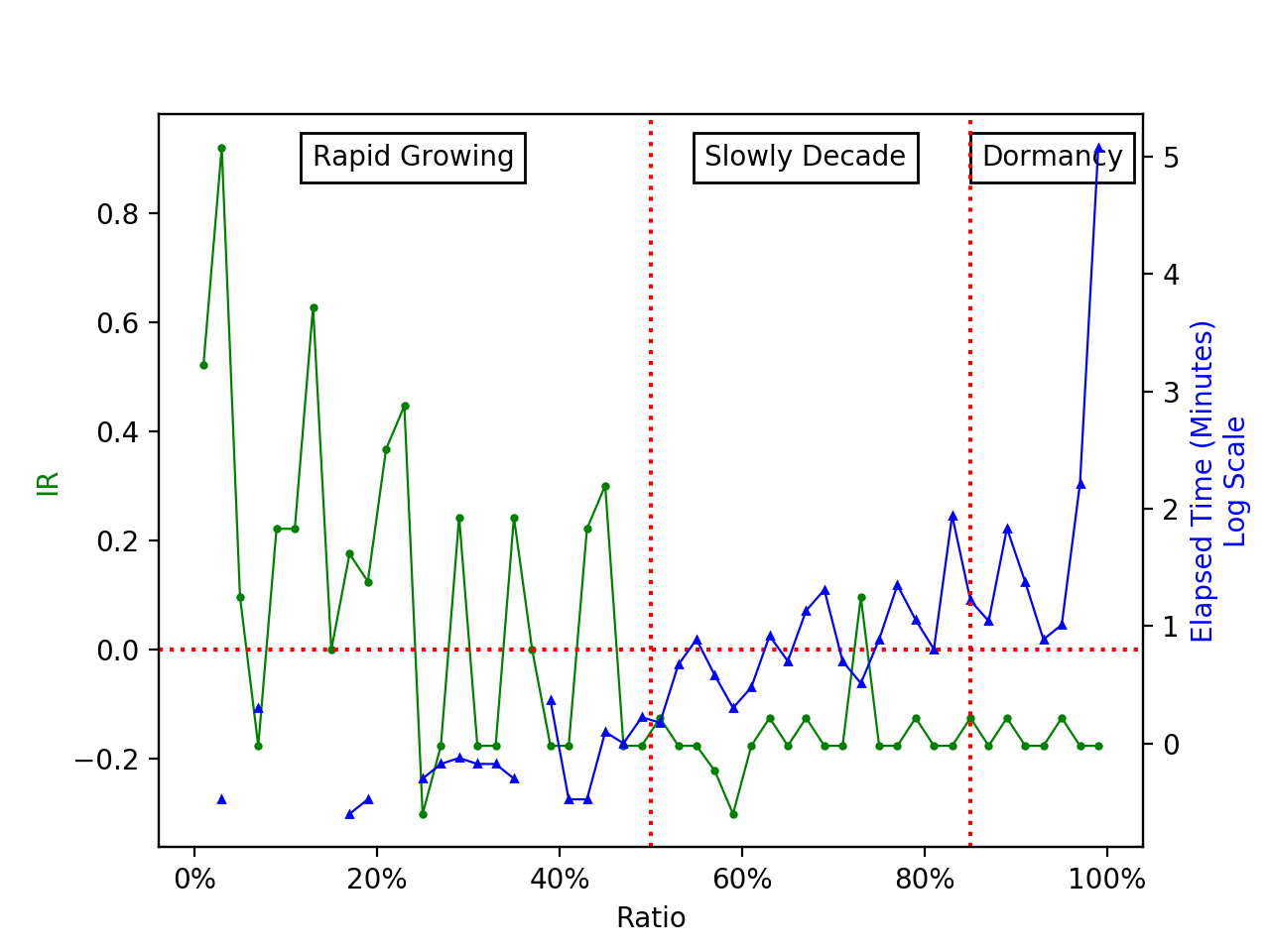}
		\caption{CNN}
	\end{subfigure}
		\begin{subfigure}[]{0.45\textwidth}
			\includegraphics[scale=0.18]{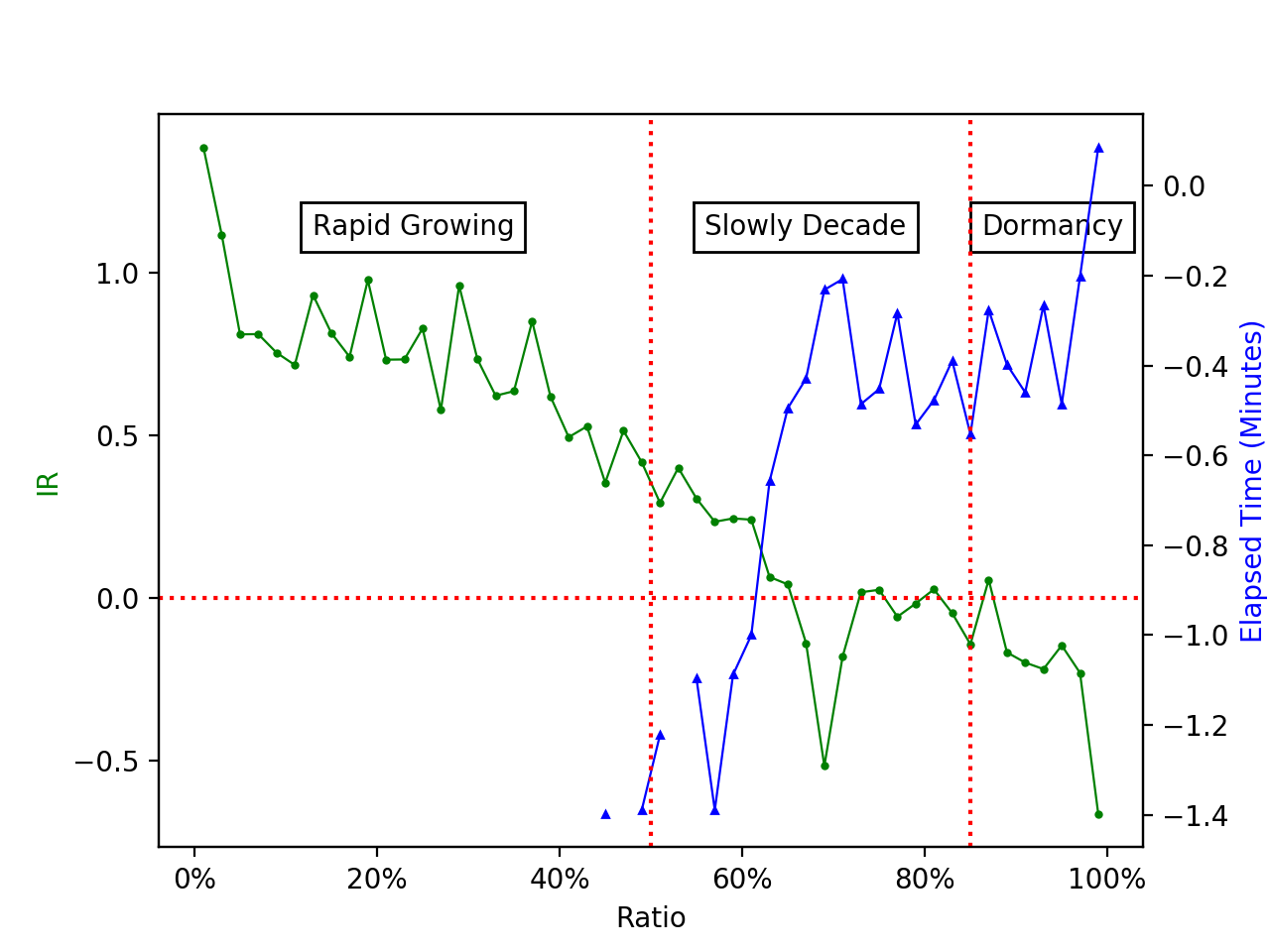}
			\caption{FOX News}
		\end{subfigure}
	\caption{Life cycle of two pages regarding IR and elapsed time}
	\label{RatioIR}
\end{figure*}

We noticed that the temporal ordering of activities on post threads are quite interesting. The audience generally rapidly increases to the peak, and then growth of the audience decays more slowly as time goes on. Figure \ref{RatioIR} visualizes the relationship among ratio, IR, and elapsed time from the last comment. We model the life cycle of post threads into three stages:
\begin{enumerate}
	\item Rapid growth: On the facebook page for FOX News, there is an obvious watershed at 50\%: From the first comment to about the midway point in post threads, many users join the discussion; usually in the next time window each comment will have 5 times the activities as the previous time window, and the time difference with last comment is usually smaller than 1 minute. However, on the CNN page, we observe that IR experiences several huge discussions, the reason may be people provide lots of reactions for some interesting comments.
	
	\item Slow Decay: For both CNN and FOX News, we observed that from 50\% to about 85\%, the volume of comments faces an obvious decline. At the same time, IR lightly decays and elapsed time becomes larger.
	\item Dormancy: At this stage, the thread has basically passed it's shelf life. The elapsed time goes up to more than ten minutes while the IR has fallen to almost 0.
\end{enumerate}

\begin{figure}[]
	\centering
	\hspace*{-0.7cm} 
	\includegraphics[scale=0.5]{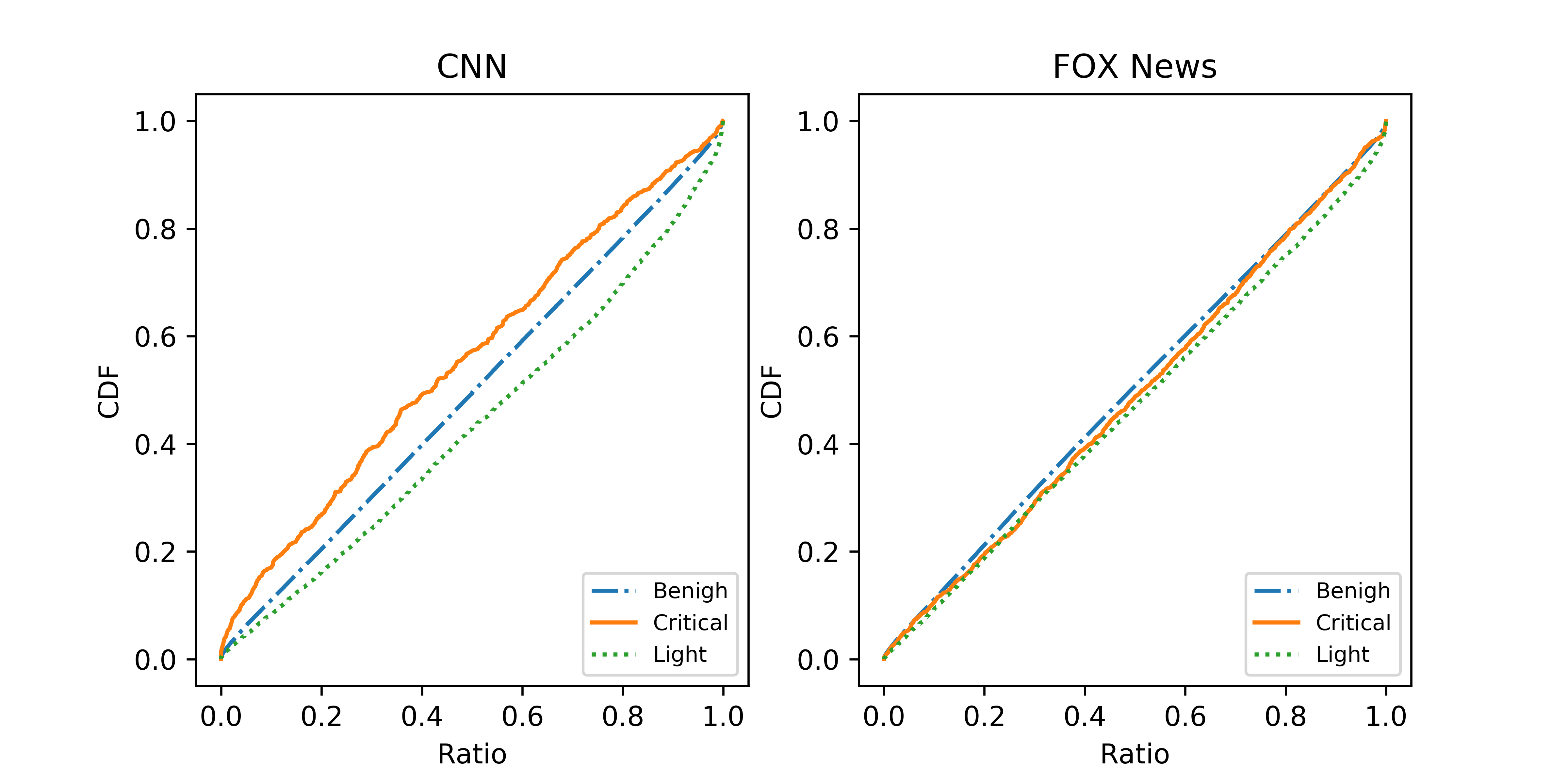}
	\caption{CDFs of different categories occurrences plotted against life stages of targeted threads}
	\label{CNNFOXRatioCDF}
\end{figure}

We also noticed that attackers are more likely to spread malicious URLs at either the Slowly Decay or Dormancy stages on CNN, while on FOX, the ratio seems to be uniform distribution, as Figure \ref{CNNFOXRatioCDF} shows.

\begin{figure*}[]
	\centering
	\begin{subfigure}[]{0.90\textwidth}
		\includegraphics[width=1\linewidth]{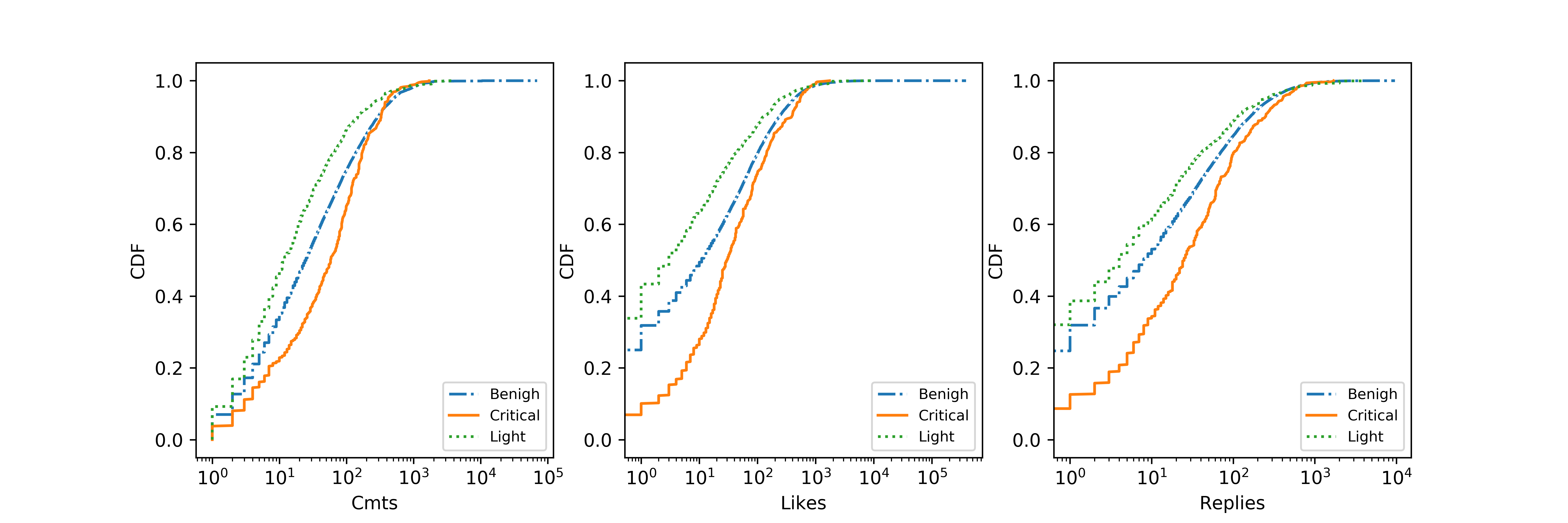}
		\caption{CNN}
	\end{subfigure}
	\begin{subfigure}[]{0.90\textwidth}
		\includegraphics[width=1\linewidth]{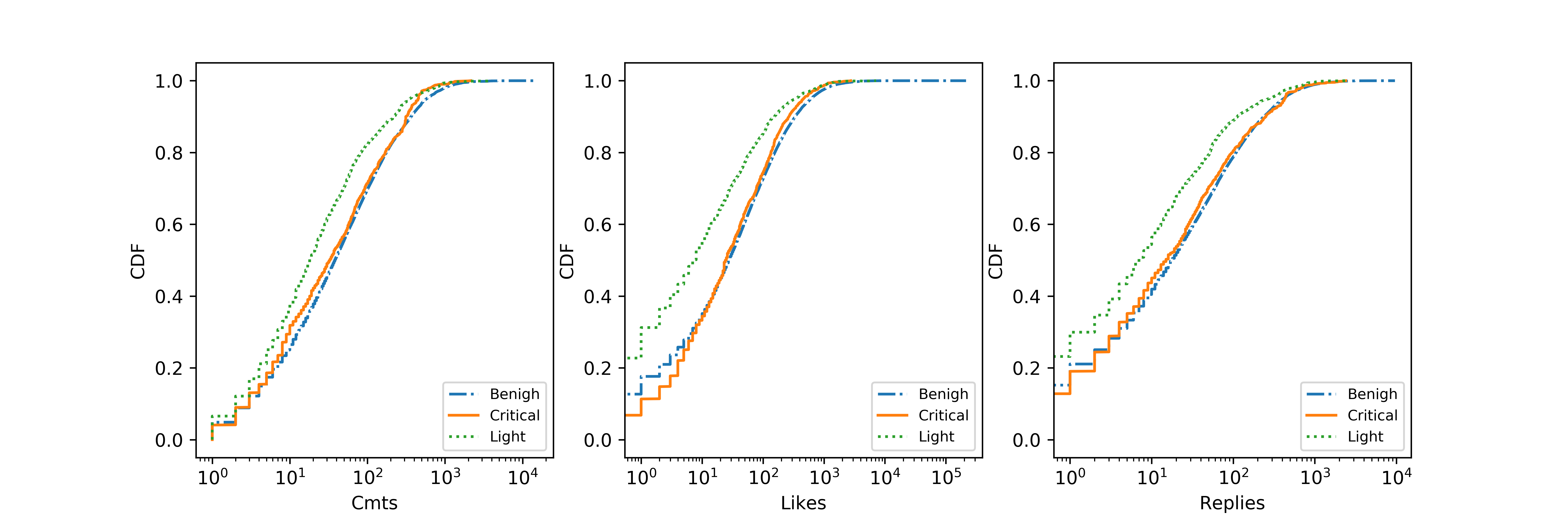}
		\caption{FOX}
	\end{subfigure}
	\caption{Users General Activities on Facebook Discussion Groups}
	\label{FOXUser}
\end{figure*}

\subsection{Attackers' footprint}
In addition to the post thread life cycle we described, we are also interested in the user's other activities. Are they actively expressing their opinions, or just acting as a one-time inappropriate information generator? We consider user activities on more than 40,000 public pages around the world from 2011 to 2016 on Facebook. 
Figure \ref{FOXUser} shows that no matter the numbers of comments, number of likes or number of reactions, those users who spread Critical-type malicious URLs occur more often than benign and Light-type users. Considered from their purpose, we noticed that Light-type users tend to lure users to commercial websites. However, Critical-type users comments usually advocate relatively personal belief and values --- which makes them heavier Facebook user and tend to influence others. However, on FOX News, only Light-type users have less activities than the other two, so there is no obvious difference between Critical-Type users and Benign Users. 

\section{Related work}
\label{section6}
Security issues surrounding OSN platforms have been growing in importance and profile due to the increasing number of users (and subsequent potential targets) of social media applications. Our related work mainly falls into two categories: (1) Popularity on social media; and (2) cyber attack techniques on social media.
\subsection{Information Diffusion and Influence}
Castillo et al. shows different categories of news (News vs. In-depth) will have different life cycles regarding social media reactions \cite{castillo2014characterizing}. In fact, there are lots of works aiming at observing and predicting the final cascade size of a given post or topic. According to Cheng et al., these factors may include content features, authors features, resharer features and temporal features which make a cascade size more predictable \cite{cheng2014can}. 
As to the last issue, several papers exploit the reaction for a given fixed time frame to predict whether a post thread will be popular or not  \cite{kupavskii2012prediction} \cite{ma2013predicting} \cite{tsur2012s}. 

Cascade can also be interpreted from the audience's perspective. I.e., why do people spend lots of time on social media to share her own opinions to public. Marwick \emph{et al.} proposed a many-to-many communication model through which individuals conceptualize an \emph{imagined audience} evoked through their content \cite{marwick2011tweet}. Hall \emph{et al.} \cite{hall2016social} demonstrates the impact of the individual on an information cascade. In the interactive communication model \cite{chapanis1975interactive}, in order to participate in the so-called attention economy, people want to attract “eyeballs”
in a media-saturated, information-rich world and to influence audiences to like their comments / photos \cite{fairchild2007building} \cite{marwick2015instafame} \cite{senft2008camgirls}. Hence, users strategically formulate their profile and participate in many discussion groups to increase attention.

\subsection{Cyber attack Analysis on Social Media}
Though increasing importance has been attached to security issues on social media, most works focus on pursuing a perfect classifier to detect malicious accounts or users with commonly used profile characteristics such as age, number of followers, geo-location and total number of activities \cite{alsaleh2014tsd} \cite{miller2014twitter}. with the development of new security threats on social media such as cyberbullying or fake news, recent research uses social science to understand collective human behavior. Cheng et al. studied the activity differences between organized groups and individuals \cite{cheng2014can}. Charzakou et al. noted people who spread hate are more engaged than typical users \cite{chatzakou2017hate}. 
Vosoughi et al. found that false news was more novel than true news mainly because of humans, not robots \cite{vosoughi2018spread}. 

\section{Conclusion}
\label{section7}
In this paper, we describe our work regarding attacker intention and influence from large-scale malicious URLs campaigns using the public Facebook Discussion Groups dataset. Specifically, we focus on examining the differing characteristics between CNN and FOX News discussion threads ranging from 2014 to 2016. 

We describe how social recommendation systems work for both target and non-target threads. Moreover, we define an Influence Ratio(IR) for every visible comment on Facebook based on the ratio between the upcoming activities and the preceding activities. We also propose a context-free prediction system to predict whether the trends will decrease or increase with a F1-score over 75\%. From these results, we perform an in-depth analysis on different categories of malicious campaigns. Compared to those comments embedded with more critical level threats such as malicious URLs, some lower level threats, such as advertising or commercial shopping URLs appeared at the very end of the discussion thread. The IR for those commercial sites for at least two reasons. (1) People just ignored those since they already know it only hinders the readability. (2) People do not want to check those posts anymore. However, the program bot did not update to the new-coming information.

The initial results we obtained provide us new insight regarding how malicious URLs influence both post thread life cycle and audience activities with the Facebook social recommendation algorithm. Our current observations enables us to reconsider new response strategies in handling inappropriate information on social media.

\bibliographystyle{plain}
\bibliography{refs}

\begin{thebibliography}{10}

\bibitem{Shalla}
Shalla secure services {Shalla's Blacklists}.
\newblock \url{http://www.shallalist.de}.
\newblock Accessed: 2018-08-27.

\bibitem{allen1965situational}
Vernon~L Allen.
\newblock Situational factors in conformity1.
\newblock In {\em Advances in experimental social psychology}, volume~2, pages
  133--175. Elsevier, 1965.

\bibitem{alsaleh2014tsd}
Mansour Alsaleh, Abdulrahman Alarifi, Abdul~Malik Al-Salman, Mohammed Alfayez,
  and Abdulmajeed Almuhaysin.
\newblock Tsd: Detecting sybil accounts in twitter.
\newblock In {\em Machine Learning and Applications (ICMLA), 2014 13th
  International Conference on}, pages 463--469. IEEE, 2014.

\bibitem{bikhchandani1992theory}
Sushil Bikhchandani, David Hirshleifer, and Ivo Welch.
\newblock A theory of fads, fashion, custom, and cultural change as
  informational cascades.
\newblock {\em Journal of political Economy}, 100(5):992--1026, 1992.

\bibitem{castillo2014characterizing}
Carlos Castillo, Mohammed El-Haddad, J{\"u}rgen Pfeffer, and Matt Stempeck.
\newblock Characterizing the life cycle of online news stories using social
  media reactions.
\newblock In {\em Proceedings of the 17th ACM conference on Computer supported
  cooperative work \& social computing}, pages 211--223. ACM, 2014.

\bibitem{chapanis1975interactive}
Alphonse Chapanis.
\newblock Interactive human communication.
\newblock {\em Scientific American}, 232(3):36--46, 1975.

\bibitem{chatzakou2017hate}
Despoina Chatzakou, Nicolas Kourtellis, Jeremy Blackburn, Emiliano
  De~Cristofaro, Gianluca Stringhini, and Athena Vakali.
\newblock Hate is not binary: Studying abusive behavior of\# gamergate on
  twitter.
\newblock In {\em Proceedings of the 28th ACM conference on hypertext and
  social media}, pages 65--74. ACM, 2017.

\bibitem{cheng2014can}
Justin Cheng, Lada Adamic, P~Alex Dow, Jon~Michael Kleinberg, and Jure
  Leskovec.
\newblock Can cascades be predicted?
\newblock In {\em Proceedings of the 23rd international conference on World
  wide web}, pages 925--936. ACM, 2014.

\bibitem{davenport2001attention}
Thomas~H Davenport and John~C Beck.
\newblock {\em The attention economy: Understanding the new currency of
  business}.
\newblock Harvard Business Press, 2001.

\bibitem{efron1994introduction}
Bradley Efron and Robert~J Tibshirani.
\newblock {\em An introduction to the bootstrap}.
\newblock CRC press, 1994.

\bibitem{fairchild2007building}
Charles Fairchild.
\newblock Building the authentic celebrity: The “idol” phenomenon in the
  attention economy.
\newblock {\em Popular Music and Society}, 30(3):355--375, 2007.

\bibitem{hall2016social}
Robert~T Hall, Joshua~S White, and Jeremy Fields.
\newblock Social relevance: toward understanding the impact of the individual
  in an information cascade.
\newblock In {\em Cyber Sensing 2016}, volume 9826, page 98260C. International
  Society for Optics and Photonics, 2016.

\bibitem{helsper2017rich}
Ellen~J Helsper and Alexander~JAM Van~Deursen.
\newblock Do the rich get digitally richer? quantity and quality of support for
  digital engagement.
\newblock {\em Information, Communication \& Society}, 20(5):700--714, 2017.

\bibitem{hong2018profiling}
Yunfeng Hong, Yu-Cheng Lin, Chun-Ming Lai, S~Felix Wu, and George~A Barnett.
\newblock Profiling facebook public page graph.
\newblock In {\em 2018 International Conference on Computing, Networking and
  Communications (ICNC)}, pages 161--165. IEEE, 2018.

\bibitem{kupavskii2012prediction}
Andrey Kupavskii, Liudmila Ostroumova, Alexey Umnov, Svyatoslav Usachev, Pavel
  Serdyukov, Gleb Gusev, and Andrey Kustarev.
\newblock Prediction of retweet cascade size over time.
\newblock In {\em Proceedings of the 21st ACM international conference on
  Information and knowledge management}, pages 2335--2338. ACM, 2012.

\bibitem{lai2017attacking}
Chun-Ming Lai, Xiaoyun Wang, Yunfeng Hong, Yu-Cheng Lin, S~Felix Wu, Patrick
  McDaniel, and Hasan Cam.
\newblock Attacking strategies and temporal analysis involving facebook
  discussion groups.
\newblock In {\em Network and Service Management (CNSM), 2017 13th
  International Conference on}, pages 1--9. IEEE, 2017.

\bibitem{lee2016predicting}
Jieun Lee and Ilyoo~B Hong.
\newblock Predicting positive user responses to social media advertising: The
  roles of emotional appeal, informativeness, and creativity.
\newblock {\em International Journal of Information Management},
  36(3):360--373, 2016.

\bibitem{ma2013predicting}
Zongyang Ma, Aixin Sun, and Gao Cong.
\newblock On predicting the popularity of newly emerging hashtags in twitter.
\newblock {\em Journal of the Association for Information Science and
  Technology}, 64(7):1399--1410, 2013.

\bibitem{marwick2015instafame}
Alice~E Marwick.
\newblock Instafame: Luxury selfies in the attention economy.
\newblock {\em Public culture}, 27(1 (75)):137--160, 2015.

\bibitem{marwick2011tweet}
Alice~E Marwick and Danah Boyd.
\newblock I tweet honestly, i tweet passionately: Twitter users, context
  collapse, and the imagined audience.
\newblock {\em New media \& society}, 13(1):114--133, 2011.

\bibitem{miller2014twitter}
Zachary Miller, Brian Dickinson, William Deitrick, Wei Hu, and Alex~Hai Wang.
\newblock Twitter spammer detection using data stream clustering.
\newblock {\em Information Sciences}, 260:64--73, 2014.

\bibitem{Pedregosa:2011:SML:1953048.2078195}
Fabian Pedregosa, Ga\"{e}l Varoquaux, Alexandre Gramfort, Vincent Michel,
  Bertrand Thirion, Olivier Grisel, Mathieu Blondel, Peter Prettenhofer, Ron
  Weiss, Vincent Dubourg, Jake Vanderplas, Alexandre Passos, David Cournapeau,
  Matthieu Brucher, Matthieu Perrot, and \'{E}douard Duchesnay.
\newblock Scikit-learn: Machine learning in python.
\newblock {\em J. Mach. Learn. Res.}, 12:2825--2830, November 2011.

\bibitem{ricci2011introduction}
Francesco Ricci, Lior Rokach, and Bracha Shapira.
\newblock Introduction to recommender systems handbook.
\newblock In {\em Recommender systems handbook}, pages 1--35. Springer, 2011.

\bibitem{senft2008camgirls}
Theresa~M Senft.
\newblock {\em Camgirls: Celebrity and community in the age of social
  networks}.
\newblock Peter Lang, 2008.

\bibitem{tsur2012s}
Oren Tsur and Ari Rappoport.
\newblock What's in a hashtag?: content based prediction of the spread of ideas
  in microblogging communities.
\newblock In {\em Proceedings of the fifth ACM international conference on Web
  search and data mining}, pages 643--652. ACM, 2012.

\bibitem{van2015off}
Bram Van~Ginneken, Arnaud~AA Setio, Colin Jacobs, and Francesco Ciompi.
\newblock Off-the-shelf convolutional neural network features for pulmonary
  nodule detection in computed tomography scans.
\newblock In {\em Biomedical Imaging (ISBI), 2015 IEEE 12th International
  Symposium on}, pages 286--289. IEEE, 2015.

\bibitem{vosoughi2018spread}
Soroush Vosoughi, Deb Roy, and Sinan Aral.
\newblock The spread of true and false news online.
\newblock {\em Science}, 359(6380):1146--1151, 2018.

\bibitem{wang2015bandwagon}
Keith~C Wang, Chun-Ming Lai, Teng Wang, and S~Felix Wu.
\newblock Bandwagon effect in facebook discussion groups.
\newblock In {\em Proceedings of the ASE BigData \& SocialInformatics 2015},
  page~17. ACM, 2015.

\end{thebibliography}
\vspace{12pt}

\end{document}